\documentclass[reprint, prl, showpacs, amsmath, amssymb, aps]{revtex4-1}
\usepackage{amsmath}
\usepackage{graphicx}
\usepackage{color}
\usepackage{bm}
\usepackage{dcolumn}

\begin{document}


\title{High field magneto-transport in two-dimensional electron gas LaAlO$_3$/SrTiO$_3$} 



\author{Ming Yang}
\email[]{ming.yang@lncmi.cnrs.fr ; phyarian@nus.edu.sg}
\affiliation{Laboratoire National des Champs Magn$\acute{e}$tiques Intenses (LNCMI-EMFL), UPR 3228, CNRS-UJF-UPS-INSA, 143 Avenue de Rangueil, 31400 Toulouse, France}

\author{Mathieu Pierre}
\affiliation{Laboratoire National des Champs Magn$\acute{e}$tiques Intenses (LNCMI-EMFL), UPR 3228, CNRS-UJF-UPS-INSA, 143 Avenue de Rangueil, 31400 Toulouse, France}

\author{Olivier Toressin}
\affiliation{Laboratoire National des Champs Magn$\acute{e}$tiques Intenses (LNCMI-EMFL), UPR 3228, CNRS-UJF-UPS-INSA, 143 Avenue de Rangueil, 31400 Toulouse, France}

\author{Michel Goiran}
\affiliation{Laboratoire National des Champs Magn$\acute{e}$tiques Intenses (LNCMI-EMFL), UPR 3228, CNRS-UJF-UPS-INSA, 143 Avenue de Rangueil, 31400 Toulouse, France}

\author{Shengwei Zeng}
\affiliation{Department of Physics and NUSNNI-Nanocore, National University of Singapore, Singapore 117411}

\author{Zhen Huang}
\affiliation{Department of Physics and NUSNNI-Nanocore, National University of Singapore, Singapore 117411}

\author{Han Kun}
\affiliation{Department of Physics and NUSNNI-Nanocore, National University of Singapore, Singapore 117411}

\author{Thirumalai Venkatesan}
\affiliation{Department of Physics and NUSNNI-Nanocore, National University of Singapore, Singapore 117411}

\author{Ariando$^*$}
\affiliation{Department of Physics and NUSNNI-Nanocore, National University of Singapore, Singapore 117411}
\email[]{phyarian@nus.edu.sg}

\author{Walter Escoffier}
\affiliation{Laboratoire National des Champs Magn$\acute{e}$tiques Intenses (LNCMI-EMFL), UPR 3228, CNRS-UJF-UPS-INSA, 143 Avenue de Rangueil, 31400 Toulouse, France}

\author{Michael Coey}
\affiliation{School of Physics and CRANN, Trinity College, Dublin, Ireland}
\affiliation{Department of Physics and NUSNNI-Nanocore, National University of Singapore, Singapore 117411}


\date{\today}

\begin{abstract}
Transport properties of the complex oxide LaAlO$_3$/SrTiO$_3$ interface are investigated under high magnetic field (55T). By rotating the sample with respect to the magnetic field, the two-dimensional nature of charge transport is clearly established. Small oscillations of the magnetoresistance with altered periodicity are observed when plotted versus inverse magnetic field. We attribute this effect to Rashba spin-orbit coupling which remains consistent with large negative magnetoresistance when the field is parallel to the sample plane. A large inconsistency between the carrier density extracted from Shubnikov-de Haas analysis and from the Hall effect is explained by the contribution to transport of at least two bands with different mobility.
\end{abstract}


\maketitle 



%
%

%


Oxide interfaces constitute a rapidly developing field of research, with potential applications in electronics \cite{mannhartoxide2010, hwangemergent2012} or solar energy harvesting \cite{assmannoxide2013}. There is currently a focus on the band-insulators LaAlO$_3$ (LAO) and SrTiO$_3$ (STO), which host a conducting two-dimensional electron gas (2DEG) at their interface \cite{ohtomoa2004}. It is mainly believed to originate from the polar catastrophe \cite{mannharttwo-dimensional2008}, which results in a charge transfer between the polar oxide [100] LAO and the nonpolar oxide [100] STO. This charge transfer prevents a divergence of the electrostatic potential associated with the intra-layer built-in electric field. Charge accumulation is therefore predicted at the interface with intriguing consequences such as magnetism \cite{brinkmanmagnetic2007} and superconductivity \cite{reyrensuperconducting2007}. In LAO/STO heterostructures, symmetry-lowering at the interface raises the Ti t$_{2g}$ band degeneracy so that the $d_{xy}$ orbital is lower in energy than the d$_{xz}$ and d$_{yz}$ orbitals. Depending on the total two-dimensional carrier density, the band occupation and the spatial distribution of the carriers \cite{huangconducting2013} are critical to envision band engineering for oxide electronics \cite{förgfield-effect2012, stornaiuolotunable2015}. Experiments utilizing a capping layer \cite{huijbendefect2013}, tuning growth temperature \cite{cavigliatwo-dimensional2010}, or using ionic liquid gating \cite{zengliquid-gated2016} have allowed for a sufficiantly high-mobility 2DEG to display Shubnikov-de Haas (SdH) oscillations under magnetic field \cite{cavigliatwo-dimensional2010, mccollamquantum2014, fetelarge2014}, opening new perspectives for the investigation of the charge carriers' properties in relation to their band-structure. However, quantum transport studies remain scarce in the literature and the large variability in the results (originating from the large range of samples studied) does not yet offer a clear picture of electron transport in LAO/STO. In this context, we make use of very large magnetic field (55 T) to extend the range of magnetoresistance measurements and enhance the visibility of SdH oscillations. We interpret our experimental data by the presence of low and high mobility electrons both contributing to transport, and confirm the role of Rashba spin-orbit coupling.

\begin{figure} [ht]
\includegraphics[width=\linewidth]{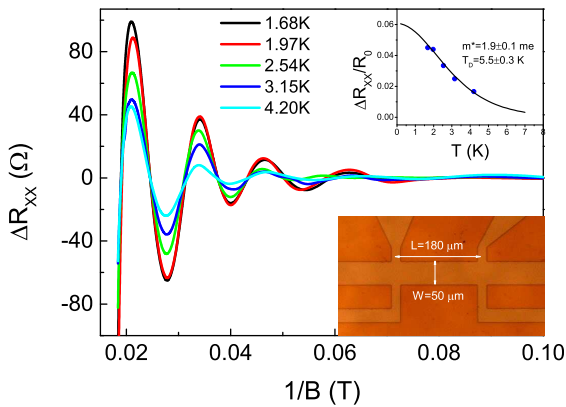}
\caption{\label{fig.1} Temperature dependence of the Shubnikov-de Haas oscillations in sample $S_1$ as a function of the inverse magnetic field after subtracting a smoothed background. Notice that the oscillation frequency is not exactly $1/B$-periodic but depends on magnetic field. The inserts show the oscillations' amplitude fitted using the Lifshitz-Kosevich expression and a photo of the sample under optical microscope.}
\end{figure}

Two samples named S$_1$ and S$_2$ were obtained by depositing 10 unit cells of LAO on TiO$_2$-terminated (100)-oriented STO substrates using pulsed laser deposition (PLD) \cite{ariandoelectronic2011}. Since both samples displayed similar results, we shall mainly discuss sample S$_1$ and make reference to sample S$_2$ only when relevant (full data for sample S$_2$ are available in the Supplemental Information). The LAO was grown at T=740$^o$C in oxygen partial pressure of $P_{O_2}=2\times10^{-3}$ Torr. During the deposition, in-situ reflection high energy electron diffraction (RHEED) was used to precisely control the layer-by-layer growth of LAO. The laser used in this work is a Lambda Physik Excimer KrF UV laser with wavelength of 248 nm at a pulse rate of 1 Hz. After deposition, the samples were cooled down to room temperature at a rate of 15$^o$C/min in the same oxygen pressure as for the deposition. The samples were then annealed in a tube furnace at 550$^o$C for 1 hour in air in order to remove the oxygen vacancies in the STO substrates introduced by high-energy plasma bombardment during the deposition \cite{liuorigin2013}. The six-terminal Hall bar devices of width W=50 $\mu$m and length L=180 $\mu$m between the longitudinal probes were fabricated by conventional photolithography using amorphous AlN films as hard masks (see insert of figure \ref{fig.1}). The devices are electrically contacted using aluminum wedge bonding. The magnetoresistance and the Hall resistance were simultaneously measured during a pulse of magnetic field of up to 55T with duration 300 ms, using a DC current {$i^{DC}=10$ $\mu$A}. Small quantum oscillations with amplitude $\sim$ 1\% of the sheet resistance value were observed on top of a large monotonic background. To improve the signal-to-noise ratio and reveal the very faint amplitude of the oscillations, another experimental run was performed where the injected current is modulated at $f=7$ kHz with RMS amplitude $i^{RMS}=10$ $\mu$A. The raw signal is later numerically demodulated and the oscillating part is extracted after subtracting a smooth background. It is worth noting that the background signal (not shown here) is strongly distorted when compared to its DC counterpart (see Supplemental Information) and is discarded from analysis. Indeed, we believe that non-ohmic contacts result in signal attenuation when measured at high frequency under pulsed magnetic field, without much affecting the oscillating part. On the other hand, the monotonic signal is reliable when measured using DC current and will be addressed later. The samples can be rotated {\it in situ} with the current aligned along the magnetic field in the parallel configuration. The angle between the rotator and the magnetic field is precisely controlled using a pick-up coil. The typical sheet resistance of device S$_1$ is 74.5 k$\Omega_\square$ at 250 K and drops down to 417 $\Omega_\square$ at 4.2 K, in line with typical values in the literature \cite{ariandoelectronic2011}.\\

We first focus on the SdH oscillations displayed in figure \ref{fig.1} and emphasise the imperfect periodicity of the oscillations when plotted as a function of inverse magnetic field. Indeed, careful experimental analysis indicates a period shift of $\sim 2$ T$^{-1}$ as the inverse magnetic field decreases. Traditionally, non-periodic oscillating features are interpreted in terms of the presence of several charge carriers, originating from different sub-bands, contributing to SdH oscillations with different periods. Usually, A Fourier transform of the raw signal usually allows extraction of the frequency peaks associated with the charge carriers. In the present case however, this procedure fails mainly because of the small number of oscillations and their {\it almost periodic} character. Furthermore, the general shape of the oscillations (the exponential-like envelope) is suggestive of a single band contribution. An alternative explanation involving spin-orbit coupling will therefore be developed. Since the effective mass $m^*$ enters in the equations of this model, we first estimate this parameter (as well as the quantum mobility $\mu_q=e\tau_q/m^*$) by studying the temperature dependance of the amplitude of the SdH oscillations using the Lifshitz-Kosevich (LK) equation below.

\begin{equation}
\Delta R_{xx}(T) = 4R_0 \times \exp\left(-\beta T_D\right)\times\frac{\beta T}{\sinh\left(\beta T\right)}
\label{eq.1}
\end{equation}

Here, $\Delta R_{xx}$ is the oscillation amplitude at a given magnetic field $B$, $R_0$ is the non-oscillatory component of the magnetoresistance, $\beta=\frac{2\pi^2 k_B m^*}{\hbar e B}$ is a prefactor while $T_D=\hbar/2\pi k_B \tau_q$ and $\tau_q$ are the Dingle temperature and the quantum mean free path, respectively. The best fit of $\Delta R_{xx}(T)$ is obtained with parameters $m^*=1.9\pm0.1\times m_e$ (where $m_e=9.1\times10^{-31}$ kg is the bare electron mass) and $\mu_q=203\pm 15$ cm$^2$/Vs (see insert of figure \ref{fig.1}). Several authors \cite{cavigliatunable2010, feterashba2012, fetelarge2014} have considered the influence of the Rashba spin-orbit coupling in the electronic properties of LAO/STO interface, arising from the interfacial breaking of inversion symmetry. In the presence of a strong perpendicular magnetic field, the usual Landau Level spectrum is modified and reads:

\begin{equation}
\left \{
\begin{array}{l l l}
E_{(N=0)}&=&\kappa_{(B)}\\
E^{\pm}_{(N>0)}&=&N \hbar\omega_c\mp\sqrt{\kappa_{(B)}^2+N \frac{2\alpha^2 e B}{\hbar}}
\end{array}
\right .\\
\label{eq.2}
\end{equation}

\begin{figure} [ht]
\includegraphics[width=\linewidth]{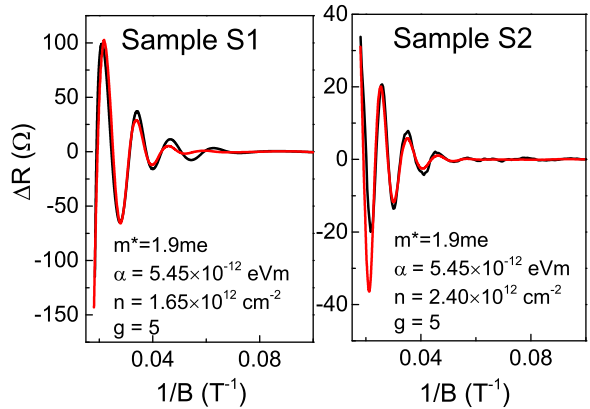}
\caption{\label{fig.2} Fit (red line) of SdH oscillations for samples $S_1$ and $S_2$ using equation \ref{eq.2} in the presence of spin-orbit coupling. The black line corresponds to the experimental data.}
\end{figure}
\begin{figure} [ht]
\includegraphics[width=\linewidth]{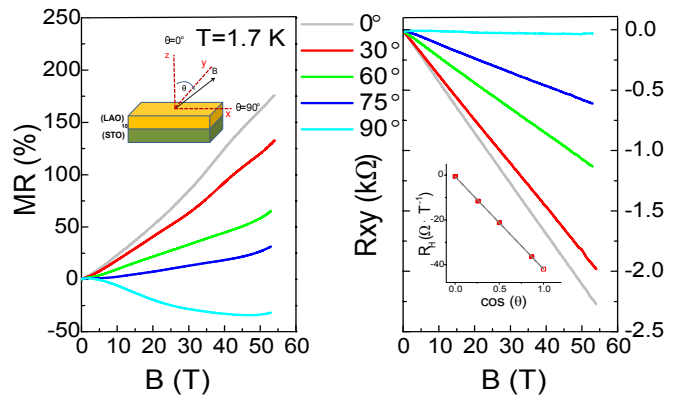}
\caption{\label{fig.3} Longitudinal magnetoresistance (a) and Hall resistance (b) at T=1.7 K as a function of the tilt angle between the sample's plane and the magnetic field for sample $S_1$. Insert of (b) displays the linear variation of the Hall coefficient versus the cosine of the tilt angle. The dash line in panel (b) is the fit of $R_{xy}(B)$ using the two-fluid model with $n_1=1.65\times10^{12}$ cm$^{-2}$, $\mu_{t,1}=1872$ cm$^2$/Vs, $n_2=1.24\times10^{13}$ cm$^{-2}$, $\mu_{t,2}=948$ cm$^2$/Vs.}
\end{figure}


where $\kappa_{(B)}=\frac{1}{2} \hbar\omega_c-\frac{1}{2} g^* \mu_B B$, $N$ is a positive integer, $\omega_c=\frac{e B}{m^*}$ is the cyclotron pulsation, $g^*$ is the spin-orbit enhanced Lande factor and $\alpha$ is the strength of the spin-orbit coupling. Following the lines of reference \cite{fetelarge2014}, the Fermi energy is computed by equating the total density of states at a given magnetic field to the fixed carrier density of the system. Considering the field-dependent orbital degeneracy of the Landau Levels (LLs) and their spectrum given by equation \ref{eq.2}, the Fermi energy actually evolves non-monotonically within the LL band-structure. As the Fermi energy alternatively crosses a LL or remains in between two LLs, the magnetoresistance oscillates respectively above or below the mean resistance value, giving rise to SdH oscillations. LL broadening is taken into account using a Gaussian line shape with $\sqrt{B}$ variance. The $B$-rising amplitude of the oscillations is finally adjusted using an exponential function. Figure \ref{fig.2} shows the best fit obtained using this procedure for sample S$_1$ and S$_2$, from which we extract the free parameters $n$, $g^*$ and $\alpha$, $m^*$ is constrained to be $1.9 m_e$ as found earlier (see insert of figure \ref{fig.1}). We would like to emphasis that the set of parameters $\{\alpha=5.45\times10^{-12}$ eVm, $g^*=5\}$ is unique for each sample provided the carrier density remains close to the one computed using the usual Onsager relation $n=(e d_s)/(h T_{SdH})$. Here, $d_s=2$ is the spin degeneracy and $T_{SdH}$ is the mean period of the oscillations plotted against inverse magnetic field (neglecting the imperfect periodicity of the SdH oscillations). The obtained carrier density $n\sim 10^{12}$ cm$^{-2}$ is of the same order of magnitude as found in other studies of similar samples \cite{shalomshubnikovde2010, zengliquid-gated2016}, but remains almost two orders of magnitude lower than the predictions of the polar catastrophe model. Furthermore, it is inconsistent with the value extracted from the linear Hall effect $n_{H}=1.48\times10^{13}$ cm$^{-2}$ shown in figure \ref{fig.3}. This discrepancy is actually a long standing issue and several interpretations have been proposed. First, the presence of valley degeneracy has been considered in \cite{shalomshubnikovde2010} and would provide a natural explanation involving a complex band-structure. However the non-integer ratio $n_{H}/n$ \footnote{For sample S1, $n_{H}=1.48\times10^{13}$ cm$^{-2}$ and $1.54\times 10^{12}$ cm$^{-2}<n<3.71\times 10^{12}$ cm$^{-2}$ depending on $B$, so that $3.98<\frac{n_{H}}{n}<9.61$. For sample S2, $n_{H}=2.31\times10^{13}$ cm$^{-2}$ and $3.13\times 10^{12}$ cm$^{-2}<n<4.97\times 10^{12}$ cm$^{-2}$ depending on $B$ so that $4.65<\frac{n_{H}}{n}<7.38$.} in the present study does not favor this hypothesis. It is worth noting, in addition, that different ratios ranging from roughly 2 to 5 have been reported in the literature \cite{cavigliatwo-dimensional2010, shalomshubnikovde2010, mccollamquantum2014, xiequantum2014} and are therefore linked to sample's growth conditions rather than to a universal band-structure characteristic. The difference between $n_{H}$ and $n$ can be reconciled assuming one or more additional conduction channel which does not contribute to SdH oscillations. In this framework, the Hall resistance can be approached using a two-fluid model, where one type of carriers are characterized by carrier density $n_{1}$ and mobility $\mu_{t,1}$ while the other type is defined by $n_2$ and $\mu_{t,2}$. We have:

\begin{eqnarray}
R_{xy}(B)&=&\frac{B}{e}\times\frac{\left(n_1\mu_{t,1}^2+n_2\mu_{t,2}^2\right)+\left(\mu_{t,1}\mu_{t,2}B\right)^2\left(n_1+n_2\right)}{\left(n_1\mu_{t,1}+n_2\mu_{t,2}\right)^2+\left(\mu_{t,1}\mu_{t,2}B\right)^2\left(n_1+n_2\right)^2}\\ \nonumber
R_{(B=0)}&=&\frac{L}{W}\times\left(e.n_1\mu_{t,1}+e.n_2\mu_{t,2}\right)^{-1}
\label{eq.3}
\end{eqnarray}

Note that $\mu_{t,i}$ ($i=1,2$) stands for the transport mobility, not the quantum mobility $\mu_q$ defined earlier. It is worth noting that the magnetic field evolution of $R_{xx}(B)$ is discarded from this analysis since a negative magnetoresistance contribution (related to spin-orbit coupling) is not captured by this simple model. Only the zero-field sample resistance $R_{(B=0)}$ is therefore considered. 
We assume that both carrier densities $n_1$ and $n_2$ relate to 2DEG. When $\mu_{t,1}\neq\mu_{t,2}$, the two carrier model yields a non-linear Hall effect, contrary to the experimental finding where the linearity of $R_{xy}(B)$ is established for the full magnetic field range [0-55 T], but the linear $B$ behaviour of the Hall resistance is progressively restored when $\mu_{t,1}/\mu_{t,2}$ approaches unity. Setting $n_1=1.65\times 10^{12}$ cm$^{-2}$, the value derived from the SdH oscillations, the best fit for the Hall resistance is obtained for $n_2=1.24\times10^{13}$ cm$^{-2}$, $\mu_{t,2}=948$ cm$^2$/Vs and $\mu_{t,1}=1872$ cm$^2$/Vs for sample $S_1$.

Based on the above analysis, we now would like to comment on the origin and transport properties of the two electron fluids. The SdH oscillations originate from heavy-mass carriers ($m^*=1.9 m_e$) which are probably derived from $d_{xz}$ and $d_{yz}$ orbitals extending deep in the STO side of the interface \cite{fetelarge2014}. These minority carriers have a density of the order of $\sim10^{12}$ cm$^{-2}$ and a fairly high mobility of a few thousands cm$^2$/Vs. Moreover, their quantum mobility is much lower than their transport mobility by roughly a factor 9, suggesting that long-range scattering is dominant \cite{coleridgesmall-angle1991}. Indeed, the quantum mobility is linked to the averaged elastic scattering time whereas the transport mobility is determined by the total scattering weighted by the scattering angle. The quantum mobility is therefore always smaller than its transport counterpart, especially when long-range scatterers are dominant and account for quasi-isotropic diffusion processes. The presence of charged O$^{2-}$ vacancies close to the interface or in the LAO layer \cite{petroviclong-range2014} provides strong support for this hypothesis, although their influence is certainly reduced by screening as the charge distribution extends deeper in the STO layer. On the other hand, we attribute the non-oscillatory part of the magnetoresistance to charge carriers lying in the lowest energy sub-band derived from the $d_{xy}$ orbitals. Such carriers are concentrated within a few unit cells of the interface between LAO and STO and certainly experience strong scattering from ionic inter-diffusion and interface reconstruction. Consequently, these charge carriers should display low mobility and are not expected to contribute to SdH oscillations. It is worth noting that even if the two fluid model captures the essence of the underlying physics, it is certainly oversimplified to account for a progressive charge distribution from the interface to deep inside the LAO layer, involving a continuous crossover from low to high mobility carriers.\\


The magnetic field dependance of the SdH oscillation frequency deserves further attention. The apparent departure from $1/B$ periodic behaviour was already observed in reference \cite{jalantwo-dimensional2010}. The authors interpreted this result as a change of carrier density with increasing magnetic field, which would be of up to 1250\% over the full magnetic field range of the present study. However this hypothesis is inconsistent with the results of the two-fluid model (see Supplemental Information) and a more natural explanation involving Rashba spin-orbit coupling is favored. Indeed, the SdH oscillations can be pretty well fitted within this framework for both samples and the extracted free parameters (namely the carrier density, $g^*$-factor and spin-orbit strength) are in line with values recently reported in a similar system \cite{fetelarge2014}. The presence of Rashba spin-orbit coupling is also consistent with the negative magnetoresistance of the order of 35\% when the magnetic field is parallel to the plane of the sample (see figure \ref{fig.3}). The interplay between electron scattering and spin-orbit coupling in the framework of the Boltzmann formalism applicable to disordered samples can yield a giant negative magnetoresistance \cite{diezgiant2015} as experimentally observed. The growth conditions certainly have a large impact on the sample's magnetoresistance response, so that a direct comparison of our data with the results of reference \cite{diezgiant2015} is impractical, however the magnitude of the magnetoresistance and the saturation field, which strongly depends on the carrier density, are in qualitative agreement. An alternative would be to attribute the negative and saturating magnetoresistance to scattering of charge carriers by localized magnetic moments. As the magnetic field increases, spin-flip scattering is progressively reduced and translates in to a decrease of resistance. According to this hypothesis, saturation of the magnetoresistance is roughly expected when the Zeeman energy $\epsilon_Z=g^*\mu_B B^*_\parallel$ matches the spin-orbit coupling energy $\epsilon_{SO}=\frac{m^*\alpha^2}{2\hbar^2}$. Using the parameters derived above, the extracted characteristic field is one order of magnitude lower than the experimental one, which invalidates the hypothesis. Furthermore, the persistence of negative magnetoresistance at elevated temperature (20 K) is not consistent with the Kondo interpretation.

To conclude, our experimental results in high magnetic field are consistent with recent published studies insofar as they support the presence of at least two conduction channels with different mobility. The high mobility carriers, with twice the bare electron mass, are located deep in the STO material but remain sensitive to charge impurities at the surface. The presence of O$^{2-}$ vacancies is at the origin of a long-range disorder, which translates into a large difference between the Drude and Dingle scattering times. The mobile electron carriers are studied through SdH oscillations with reproducible deviations from $1/B$ periodicity. This effect is interpreted as a consequence of Rashba spin-orbit coupling, consistent with the large negative, saturating magnetoresistance when the field is applied parallel to the sample plane. On the other hand, the low mobility carriers are located close the LAO/STO interface and experience strong scattering, so that the corresponding SdH oscillations remain out of experimental reach even in magnetic fields as high as 55 T. Their contribution is visible in the linear Hall effect, with carrier density roughly one order of magnitude higher than the mobile electrons. In order to validate these conclusions, a higher magnetic field study with varying carrier concentration would be required. It will be part of our future study.

\begin{acknowledgments}
Sample fabrication by PLD was carried out at National University of Singapore. Low temperature and high magnetic field measurements were performed at LNCMI under the EMFL proposal TSC16-213. This work is partly supported by the scholarship from China Scholarship Council (CSC) under the Grant CSC N°201404490072 and the National University of Singapore (NUS) Academic Research Fund (AcRF Tier 1 Grant No. R-144-000-346-112 and R-144-000-364-112).

\end{acknowledgments}



\begin{thebibliography}{26}%
\makeatletter
\providecommand \@ifxundefined [1]{%
 \@ifx{#1\undefined}
}%
\providecommand \@ifnum [1]{%
 \ifnum #1\expandafter \@firstoftwo
 \else \expandafter \@secondoftwo
 \fi
}%
\providecommand \@ifx [1]{%
 \ifx #1\expandafter \@firstoftwo
 \else \expandafter \@secondoftwo
 \fi
}%
\providecommand \natexlab [1]{#1}%
\providecommand \enquote  [1]{``#1''}%
\providecommand \bibnamefont  [1]{#1}%
\providecommand \bibfnamefont [1]{#1}%
\providecommand \citenamefont [1]{#1}%
\providecommand \href@noop [0]{\@secondoftwo}%
\providecommand \href [0]{\begingroup \@sanitize@url \@href}%
\providecommand \@href[1]{\@@startlink{#1}\@@href}%
\providecommand \@@href[1]{\endgroup#1\@@endlink}%
\providecommand \@sanitize@url [0]{\catcode `\\12\catcode `\$12\catcode
  `\&12\catcode `\#12\catcode `\^12\catcode `\_12\catcode `\%12\relax}%
\providecommand \@@startlink[1]{}%
\providecommand \@@endlink[0]{}%
\providecommand \url  [0]{\begingroup\@sanitize@url \@url }%
\providecommand \@url [1]{\endgroup\@href {#1}{\urlprefix }}%
\providecommand \urlprefix  [0]{URL }%
\providecommand \Eprint [0]{\href }%
\providecommand \doibase [0]{http://dx.doi.org/}%
\providecommand \selectlanguage [0]{\@gobble}%
\providecommand \bibinfo  [0]{\@secondoftwo}%
\providecommand \bibfield  [0]{\@secondoftwo}%
\providecommand \translation [1]{[#1]}%
\providecommand \BibitemOpen [0]{}%
\providecommand \bibitemStop [0]{}%
\providecommand \bibitemNoStop [0]{.\EOS\space}%
\providecommand \EOS [0]{\spacefactor3000\relax}%
\providecommand \BibitemShut  [1]{\csname bibitem#1\endcsname}%
\let\auto@bib@innerbib\@empty
\bibitem [{\citenamefont {Mannhart}\ and\ \citenamefont
  {Schlom}(2010)}]{mannhartoxide2010}%
  \BibitemOpen
  \bibfield  {author} {\bibinfo {author} {\bibfnamefont {J.}~\bibnamefont
  {Mannhart}}\ and\ \bibinfo {author} {\bibfnamefont {D.}~\bibnamefont
  {Schlom}},\ }\href {\doibase 10.1126/science.1181862} {\bibfield  {journal}
  {\bibinfo  {journal} {Science {(New} York, {N.Y.)}}\ }\textbf {\bibinfo
  {volume} {327}},\ \bibinfo {pages} {1607} (\bibinfo {year}
  {2010})}\BibitemShut {NoStop}%
\bibitem [{\citenamefont {Hwang}\ \emph {et~al.}(2012)\citenamefont {Hwang},
  \citenamefont {Iwasa}, \citenamefont {Kawasaki}, \citenamefont {Keimer},
  \citenamefont {Nagaosa},\ and\ \citenamefont {Tokura}}]{hwangemergent2012}%
  \BibitemOpen
  \bibfield  {author} {\bibinfo {author} {\bibfnamefont {H.}~\bibnamefont
  {Hwang}}, \bibinfo {author} {\bibfnamefont {Y.}~\bibnamefont {Iwasa}},
  \bibinfo {author} {\bibfnamefont {M.}~\bibnamefont {Kawasaki}}, \bibinfo
  {author} {\bibfnamefont {B.}~\bibnamefont {Keimer}}, \bibinfo {author}
  {\bibfnamefont {N.}~\bibnamefont {Nagaosa}}, \ and\ \bibinfo {author}
  {\bibfnamefont {Y.}~\bibnamefont {Tokura}},\ }\href {\doibase
  10.1038/nmat3223} {\bibfield  {journal} {\bibinfo  {journal} {Nature
  materials}\ }\textbf {\bibinfo {volume} {11}},\ \bibinfo {pages} {103}
  (\bibinfo {year} {2012})}\BibitemShut {NoStop}%
\bibitem [{\citenamefont {Assmann}\ \emph {et~al.}(2013)\citenamefont
  {Assmann}, \citenamefont {Blaha}, \citenamefont {Laskowski}, \citenamefont
  {Held}, \citenamefont {Okamoto},\ and\ \citenamefont
  {Sangiovanni}}]{assmannoxide2013}%
  \BibitemOpen
  \bibfield  {author} {\bibinfo {author} {\bibfnamefont {E.}~\bibnamefont
  {Assmann}}, \bibinfo {author} {\bibfnamefont {P.}~\bibnamefont {Blaha}},
  \bibinfo {author} {\bibfnamefont {R.}~\bibnamefont {Laskowski}}, \bibinfo
  {author} {\bibfnamefont {K.}~\bibnamefont {Held}}, \bibinfo {author}
  {\bibfnamefont {S.}~\bibnamefont {Okamoto}}, \ and\ \bibinfo {author}
  {\bibfnamefont {G.}~\bibnamefont {Sangiovanni}},\ }\href {\doibase
  10.1103/PhysRevLett.110.078701} {\bibfield  {journal} {\bibinfo  {journal}
  {Physical review letters}\ }\textbf {\bibinfo {volume} {110}},\ \bibinfo
  {pages} {078701} (\bibinfo {year} {2013})}\BibitemShut {NoStop}%
\bibitem [{\citenamefont {Ohtomo}\ and\ \citenamefont
  {Hwang}(2004)}]{ohtomoa2004}%
  \BibitemOpen
  \bibfield  {author} {\bibinfo {author} {\bibfnamefont {A.}~\bibnamefont
  {Ohtomo}}\ and\ \bibinfo {author} {\bibfnamefont {H.}~\bibnamefont {Hwang}},\
  }\href {\doibase 10.1038/nature02308} {\bibfield  {journal} {\bibinfo
  {journal} {Nature}\ }\textbf {\bibinfo {volume} {427}},\ \bibinfo {pages}
  {423} (\bibinfo {year} {2004})}\BibitemShut {NoStop}%
\bibitem [{\citenamefont {Mannhart}\ \emph {et~al.}(2008)\citenamefont
  {Mannhart}, \citenamefont {Blank}, \citenamefont {Hwang}, \citenamefont
  {Millis},\ and\ \citenamefont {Triscone}}]{mannharttwo-dimensional2008}%
  \BibitemOpen
  \bibfield  {author} {\bibinfo {author} {\bibfnamefont {J.}~\bibnamefont
  {Mannhart}}, \bibinfo {author} {\bibfnamefont {D.}~\bibnamefont {Blank}},
  \bibinfo {author} {\bibfnamefont {H.}~\bibnamefont {Hwang}}, \bibinfo
  {author} {\bibfnamefont {A.}~\bibnamefont {Millis}}, \ and\ \bibinfo {author}
  {\bibfnamefont {J.}~\bibnamefont {Triscone}},\ }\href {\doibase
  10.1557/mrs2008.222} {\bibfield  {journal} {\bibinfo  {journal} {{MRS}
  Bulletin}\ }\textbf {\bibinfo {volume} {33}},\ \bibinfo {pages} {1027}
  (\bibinfo {year} {2008})}\BibitemShut {NoStop}%
\bibitem [{\citenamefont {Brinkman}\ \emph {et~al.}(2007)\citenamefont
  {Brinkman}, \citenamefont {Huijben}, \citenamefont {van Zalk}, \citenamefont
  {Huijben}, \citenamefont {Zeitler}, \citenamefont {Maan}, \citenamefont
  {van~der Wiel}, \citenamefont {Rijnders}, \citenamefont {Blank},\ and\
  \citenamefont {Hilgenkamp}}]{brinkmanmagnetic2007}%
  \BibitemOpen
  \bibfield  {author} {\bibinfo {author} {\bibfnamefont {A.}~\bibnamefont
  {Brinkman}}, \bibinfo {author} {\bibfnamefont {M.}~\bibnamefont {Huijben}},
  \bibinfo {author} {\bibfnamefont {M.}~\bibnamefont {van Zalk}}, \bibinfo
  {author} {\bibfnamefont {J.}~\bibnamefont {Huijben}}, \bibinfo {author}
  {\bibfnamefont {U.}~\bibnamefont {Zeitler}}, \bibinfo {author} {\bibfnamefont
  {J.}~\bibnamefont {Maan}}, \bibinfo {author} {\bibfnamefont {W.}~\bibnamefont
  {van~der Wiel}}, \bibinfo {author} {\bibfnamefont {G.}~\bibnamefont
  {Rijnders}}, \bibinfo {author} {\bibfnamefont {D.}~\bibnamefont {Blank}}, \
  and\ \bibinfo {author} {\bibfnamefont {H.}~\bibnamefont {Hilgenkamp}},\
  }\href {\doibase 10.1038/nmat1931} {\bibfield  {journal} {\bibinfo  {journal}
  {Nature materials}\ }\textbf {\bibinfo {volume} {6}},\ \bibinfo {pages} {493}
  (\bibinfo {year} {2007})}\BibitemShut {NoStop}%
\bibitem [{\citenamefont {Reyren}\ \emph {et~al.}(2007)\citenamefont {Reyren},
  \citenamefont {Thiel}, \citenamefont {Caviglia}, \citenamefont {Kourkoutis},
  \citenamefont {Hammerl}, \citenamefont {Richter}, \citenamefont {Schneider},
  \citenamefont {Kopp}, \citenamefont {R{{\"u}}etschi}, \citenamefont
  {Jaccard}, \citenamefont {Gabay}, \citenamefont {Muller}, \citenamefont
  {Triscone},\ and\ \citenamefont {Mannhart}}]{reyrensuperconducting2007}%
  \BibitemOpen
  \bibfield  {author} {\bibinfo {author} {\bibfnamefont {N.}~\bibnamefont
  {Reyren}}, \bibinfo {author} {\bibfnamefont {S.}~\bibnamefont {Thiel}},
  \bibinfo {author} {\bibfnamefont {A.}~\bibnamefont {Caviglia}}, \bibinfo
  {author} {\bibfnamefont {L.}~\bibnamefont {Kourkoutis}}, \bibinfo {author}
  {\bibfnamefont {G.}~\bibnamefont {Hammerl}}, \bibinfo {author} {\bibfnamefont
  {C.}~\bibnamefont {Richter}}, \bibinfo {author} {\bibfnamefont
  {C.}~\bibnamefont {Schneider}}, \bibinfo {author} {\bibfnamefont
  {T.}~\bibnamefont {Kopp}}, \bibinfo {author} {\bibfnamefont {A.~S.}\
  \bibnamefont {R{{\"u}}etschi}}, \bibinfo {author} {\bibfnamefont
  {D.}~\bibnamefont {Jaccard}}, \bibinfo {author} {\bibfnamefont
  {M.}~\bibnamefont {Gabay}}, \bibinfo {author} {\bibfnamefont
  {D.}~\bibnamefont {Muller}}, \bibinfo {author} {\bibfnamefont {J.~M.}\
  \bibnamefont {Triscone}}, \ and\ \bibinfo {author} {\bibfnamefont
  {J.}~\bibnamefont {Mannhart}},\ }\href {\doibase 10.1126/science.1146006}
  {\bibfield  {journal} {\bibinfo  {journal} {Science {(New} York, {N.Y.)}}\
  }\textbf {\bibinfo {volume} {317}},\ \bibinfo {pages} {1196} (\bibinfo {year}
  {2007})}\BibitemShut {NoStop}%
\bibitem [{\citenamefont {Huang}\ \emph {et~al.}(2013)\citenamefont {Huang},
  \citenamefont {Wang}, \citenamefont {Liu}, \citenamefont {L{{\"u}}},
  \citenamefont {Zeng}, \citenamefont {Annadi}, \citenamefont {Tan},
  \citenamefont {Qiu}, \citenamefont {Zhao}, \citenamefont {Salluzzo},
  \citenamefont {Coey}, \citenamefont {Venkatesan},\ and\ \citenamefont
  {Ariando}}]{huangconducting2013}%
  \BibitemOpen
  \bibfield  {author} {\bibinfo {author} {\bibfnamefont {Z.}~\bibnamefont
  {Huang}}, \bibinfo {author} {\bibfnamefont {X.~R.}\ \bibnamefont {Wang}},
  \bibinfo {author} {\bibfnamefont {Z.~Q.}\ \bibnamefont {Liu}}, \bibinfo
  {author} {\bibfnamefont {W.~M.}\ \bibnamefont {L{{\"u}}}}, \bibinfo {author}
  {\bibfnamefont {S.~W.}\ \bibnamefont {Zeng}}, \bibinfo {author}
  {\bibfnamefont {A.}~\bibnamefont {Annadi}}, \bibinfo {author} {\bibfnamefont
  {W.~L.}\ \bibnamefont {Tan}}, \bibinfo {author} {\bibfnamefont {X.~P.}\
  \bibnamefont {Qiu}}, \bibinfo {author} {\bibfnamefont {Y.~L.}\ \bibnamefont
  {Zhao}}, \bibinfo {author} {\bibfnamefont {M.}~\bibnamefont {Salluzzo}},
  \bibinfo {author} {\bibfnamefont {J.~M.~D.}\ \bibnamefont {Coey}}, \bibinfo
  {author} {\bibfnamefont {T.}~\bibnamefont {Venkatesan}}, \ and\ \bibinfo
  {author} {\bibnamefont {Ariando}},\ }\href {\doibase
  10.1103/PhysRevB.88.161107} {\bibfield  {journal} {\bibinfo  {journal}
  {Physical Review B}\ }\textbf {\bibinfo {volume} {88}} (\bibinfo {year}
  {2013}),\ 10.1103/PhysRevB.88.161107}\BibitemShut {NoStop}%
\bibitem [{\citenamefont {Fo�rg}\ \emph {et~al.}(2012)\citenamefont {Fo�rg},
  \citenamefont {Richter},\ and\ \citenamefont
  {Mannhart}}]{fo�rgfield-effect2012}%
  \BibitemOpen
  \bibfield  {author} {\bibinfo {author} {\bibfnamefont {B.}~\bibnamefont
  {Fo�rg}}, \bibinfo {author} {\bibfnamefont {C.}~\bibnamefont {Richter}}, \
  and\ \bibinfo {author} {\bibfnamefont {J.}~\bibnamefont {Mannhart}},\ }\href
  {\doibase 10.1063/1.3682102} {\bibfield  {journal} {\bibinfo  {journal}
  {Applied Physics Letters}\ }\textbf {\bibinfo {volume} {100}},\ \bibinfo
  {pages} {053506} (\bibinfo {year} {2012})}\BibitemShut {NoStop}%
\bibitem [{\citenamefont {Stornaiuolo}\ \emph {et~al.}(2015)\citenamefont
  {Stornaiuolo}, \citenamefont {Cantoni}, \citenamefont {Luca}, \citenamefont
  {Capua}, \citenamefont {Gennaro}, \citenamefont {Ghiringhelli}, \citenamefont
  {Jouault}, \citenamefont {Marr{\`e}}, \citenamefont {Massarotti},
  \citenamefont {Granozio}, \citenamefont {Pallecchi}, \citenamefont
  {Piamonteze}, \citenamefont {Rusponi}, \citenamefont {Tafuri},\ and\
  \citenamefont {Salluzzo}}]{stornaiuolotunable2015}%
  \BibitemOpen
  \bibfield  {author} {\bibinfo {author} {\bibfnamefont {D.}~\bibnamefont
  {Stornaiuolo}}, \bibinfo {author} {\bibfnamefont {C.}~\bibnamefont
  {Cantoni}}, \bibinfo {author} {\bibfnamefont {G.~M.~D.}\ \bibnamefont
  {Luca}}, \bibinfo {author} {\bibfnamefont {R.~D.}\ \bibnamefont {Capua}},
  \bibinfo {author} {\bibfnamefont {E.~D.}\ \bibnamefont {Gennaro}}, \bibinfo
  {author} {\bibfnamefont {G.}~\bibnamefont {Ghiringhelli}}, \bibinfo {author}
  {\bibfnamefont {B.}~\bibnamefont {Jouault}}, \bibinfo {author} {\bibfnamefont
  {D.}~\bibnamefont {Marr{\`e}}}, \bibinfo {author} {\bibfnamefont
  {D.}~\bibnamefont {Massarotti}}, \bibinfo {author} {\bibfnamefont {F.~M.}\
  \bibnamefont {Granozio}}, \bibinfo {author} {\bibfnamefont {I.}~\bibnamefont
  {Pallecchi}}, \bibinfo {author} {\bibfnamefont {C.}~\bibnamefont
  {Piamonteze}}, \bibinfo {author} {\bibfnamefont {S.}~\bibnamefont {Rusponi}},
  \bibinfo {author} {\bibfnamefont {F.}~\bibnamefont {Tafuri}}, \ and\ \bibinfo
  {author} {\bibfnamefont {M.}~\bibnamefont {Salluzzo}},\ }\href {\doibase
  10.1038/nmat4491} {\bibfield  {journal} {\bibinfo  {journal} {Nature
  Materials}\ } (\bibinfo {year} {2015}),\ 10.1038/nmat4491}\BibitemShut
  {NoStop}%
\bibitem [{\citenamefont {Huijben}\ \emph {et~al.}(2013)\citenamefont
  {Huijben}, \citenamefont {Koster}, \citenamefont {Kruize}, \citenamefont
  {Wenderich}, \citenamefont {Verbeeck}, \citenamefont {Bals}, \citenamefont
  {Slooten}, \citenamefont {Shi}, \citenamefont {Molegraaf}, \citenamefont
  {Kleibeuker}, \citenamefont {van Aert}, \citenamefont {Goedkoop},
  \citenamefont {Brinkman}, \citenamefont {Blank}, \citenamefont {Golden},
  \citenamefont {van Tendeloo}, \citenamefont {Hilgenkamp},\ and\ \citenamefont
  {Rijnders}}]{huijbendefect2013}%
  \BibitemOpen
  \bibfield  {author} {\bibinfo {author} {\bibfnamefont {M.}~\bibnamefont
  {Huijben}}, \bibinfo {author} {\bibfnamefont {G.}~\bibnamefont {Koster}},
  \bibinfo {author} {\bibfnamefont {M.~K.}\ \bibnamefont {Kruize}}, \bibinfo
  {author} {\bibfnamefont {S.}~\bibnamefont {Wenderich}}, \bibinfo {author}
  {\bibfnamefont {J.}~\bibnamefont {Verbeeck}}, \bibinfo {author}
  {\bibfnamefont {S.}~\bibnamefont {Bals}}, \bibinfo {author} {\bibfnamefont
  {E.}~\bibnamefont {Slooten}}, \bibinfo {author} {\bibfnamefont
  {B.}~\bibnamefont {Shi}}, \bibinfo {author} {\bibfnamefont {H.~J.~A.}\
  \bibnamefont {Molegraaf}}, \bibinfo {author} {\bibfnamefont {J.~E.}\
  \bibnamefont {Kleibeuker}}, \bibinfo {author} {\bibfnamefont
  {S.}~\bibnamefont {van Aert}}, \bibinfo {author} {\bibfnamefont {J.~B.}\
  \bibnamefont {Goedkoop}}, \bibinfo {author} {\bibfnamefont {A.}~\bibnamefont
  {Brinkman}}, \bibinfo {author} {\bibfnamefont {D.~H.~A.}\ \bibnamefont
  {Blank}}, \bibinfo {author} {\bibfnamefont {M.~S.}\ \bibnamefont {Golden}},
  \bibinfo {author} {\bibfnamefont {G.}~\bibnamefont {van Tendeloo}}, \bibinfo
  {author} {\bibfnamefont {H.}~\bibnamefont {Hilgenkamp}}, \ and\ \bibinfo
  {author} {\bibfnamefont {G.}~\bibnamefont {Rijnders}},\ }\href {\doibase
  10.1002/adfm.201203355} {\bibfield  {journal} {\bibinfo  {journal} {Advanced
  Functional Materials}\ }\textbf {\bibinfo {volume} {23}},\ \bibinfo {pages}
  {5240} (\bibinfo {year} {2013})}\BibitemShut {NoStop}%
\bibitem [{\citenamefont {Caviglia}\ \emph
  {et~al.}(2010{\natexlab{a}})\citenamefont {Caviglia}, \citenamefont
  {Gariglio}, \citenamefont {Cancellieri}, \citenamefont {Sac{\'e}p{\'e}},
  \citenamefont {F{\^e}te}, \citenamefont {Reyren}, \citenamefont {Gabay},
  \citenamefont {Morpurgo},\ and\ \citenamefont
  {Triscone}}]{cavigliatwo-dimensional2010}%
  \BibitemOpen
  \bibfield  {author} {\bibinfo {author} {\bibfnamefont {A.~D.}\ \bibnamefont
  {Caviglia}}, \bibinfo {author} {\bibfnamefont {S.}~\bibnamefont {Gariglio}},
  \bibinfo {author} {\bibfnamefont {C.}~\bibnamefont {Cancellieri}}, \bibinfo
  {author} {\bibfnamefont {B.}~\bibnamefont {Sac{\'e}p{\'e}}}, \bibinfo
  {author} {\bibfnamefont {A.}~\bibnamefont {F{\^e}te}}, \bibinfo {author}
  {\bibfnamefont {N.}~\bibnamefont {Reyren}}, \bibinfo {author} {\bibfnamefont
  {M.}~\bibnamefont {Gabay}}, \bibinfo {author} {\bibfnamefont {A.~F.}\
  \bibnamefont {Morpurgo}}, \ and\ \bibinfo {author} {\bibfnamefont
  {J.}~\bibnamefont {Triscone}},\ }\href {\doibase
  10.1103/PhysRevLett.105.236802} {\bibfield  {journal} {\bibinfo  {journal}
  {Physical review letters}\ }\textbf {\bibinfo {volume} {105}},\ \bibinfo
  {pages} {236802} (\bibinfo {year} {2010}{\natexlab{a}})}\BibitemShut
  {NoStop}%
\bibitem [{\citenamefont {Zeng}\ \emph {et~al.}(2016)\citenamefont {Zeng},
  \citenamefont {L{{\"u}}}, \citenamefont {Huang}, \citenamefont {Liu},
  \citenamefont {Han}, \citenamefont {Gopinadhan}, \citenamefont {Li},
  \citenamefont {Guo}, \citenamefont {Zhou}, \citenamefont {Ma}, \citenamefont
  {Jian}, \citenamefont {Venkatesan},\ and\ \citenamefont
  {Ariando}}]{zengliquid-gated2016}%
  \BibitemOpen
  \bibfield  {author} {\bibinfo {author} {\bibfnamefont {S.}~\bibnamefont
  {Zeng}}, \bibinfo {author} {\bibfnamefont {W.}~\bibnamefont {L{{\"u}}}},
  \bibinfo {author} {\bibfnamefont {Z.}~\bibnamefont {Huang}}, \bibinfo
  {author} {\bibfnamefont {Z.}~\bibnamefont {Liu}}, \bibinfo {author}
  {\bibfnamefont {K.}~\bibnamefont {Han}}, \bibinfo {author} {\bibfnamefont
  {K.}~\bibnamefont {Gopinadhan}}, \bibinfo {author} {\bibfnamefont
  {C.}~\bibnamefont {Li}}, \bibinfo {author} {\bibfnamefont {R.}~\bibnamefont
  {Guo}}, \bibinfo {author} {\bibfnamefont {W.}~\bibnamefont {Zhou}}, \bibinfo
  {author} {\bibfnamefont {H.~H.}\ \bibnamefont {Ma}}, \bibinfo {author}
  {\bibfnamefont {L.}~\bibnamefont {Jian}}, \bibinfo {author} {\bibfnamefont
  {T.}~\bibnamefont {Venkatesan}}, \ and\ \bibinfo {author} {\bibnamefont
  {Ariando}},\ }\href {\doibase 10.1021/acsnano.6b00409} {\bibfield  {journal}
  {\bibinfo  {journal} {{ACS} nano}\ } (\bibinfo {year} {2016}),\
  10.1021/acsnano.6b00409}\BibitemShut {NoStop}%
\bibitem [{\citenamefont {{McCollam}}\ \emph {et~al.}(2014)\citenamefont
  {{McCollam}}, \citenamefont {Wenderich}, \citenamefont {Kruize},
  \citenamefont {Guduru}, \citenamefont {Molegraaf}, \citenamefont {Huijben},
  \citenamefont {Koster}, \citenamefont {Blank}, \citenamefont {Rijnders},
  \citenamefont {Brinkman}, \citenamefont {Hilgenkamp}, \citenamefont
  {Zeitler},\ and\ \citenamefont {Maan}}]{mccollamquantum2014}%
  \BibitemOpen
  \bibfield  {author} {\bibinfo {author} {\bibfnamefont {A.}~\bibnamefont
  {{McCollam}}}, \bibinfo {author} {\bibfnamefont {S.}~\bibnamefont
  {Wenderich}}, \bibinfo {author} {\bibfnamefont {M.~K.}\ \bibnamefont
  {Kruize}}, \bibinfo {author} {\bibfnamefont {V.~K.}\ \bibnamefont {Guduru}},
  \bibinfo {author} {\bibfnamefont {H.~J.~A.}\ \bibnamefont {Molegraaf}},
  \bibinfo {author} {\bibfnamefont {M.}~\bibnamefont {Huijben}}, \bibinfo
  {author} {\bibfnamefont {G.}~\bibnamefont {Koster}}, \bibinfo {author}
  {\bibfnamefont {D.~H.~A.}\ \bibnamefont {Blank}}, \bibinfo {author}
  {\bibfnamefont {G.}~\bibnamefont {Rijnders}}, \bibinfo {author}
  {\bibfnamefont {A.}~\bibnamefont {Brinkman}}, \bibinfo {author}
  {\bibfnamefont {H.}~\bibnamefont {Hilgenkamp}}, \bibinfo {author}
  {\bibfnamefont {U.}~\bibnamefont {Zeitler}}, \ and\ \bibinfo {author}
  {\bibfnamefont {J.~C.}\ \bibnamefont {Maan}},\ }\href {\doibase
  10.1063/1.4863786} {\bibfield  {journal} {\bibinfo  {journal} {{APL}
  Materials}\ }\textbf {\bibinfo {volume} {2}},\ \bibinfo {pages} {022102}
  (\bibinfo {year} {2014})}\BibitemShut {NoStop}%
\bibitem [{\citenamefont {F{\^e}te}\ \emph {et~al.}(2014)\citenamefont
  {F{\^e}te}, \citenamefont {Gariglio}, \citenamefont {Berthod}, \citenamefont
  {Li}, \citenamefont {Stornaiuolo}, \citenamefont {Gabay},\ and\ \citenamefont
  {Triscone}}]{fetelarge2014}%
  \BibitemOpen
  \bibfield  {author} {\bibinfo {author} {\bibfnamefont {A.}~\bibnamefont
  {F{\^e}te}}, \bibinfo {author} {\bibfnamefont {S.}~\bibnamefont {Gariglio}},
  \bibinfo {author} {\bibfnamefont {C.}~\bibnamefont {Berthod}}, \bibinfo
  {author} {\bibfnamefont {D.}~\bibnamefont {Li}}, \bibinfo {author}
  {\bibfnamefont {D.}~\bibnamefont {Stornaiuolo}}, \bibinfo {author}
  {\bibfnamefont {M.}~\bibnamefont {Gabay}}, \ and\ \bibinfo {author}
  {\bibfnamefont {J.}~\bibnamefont {Triscone}},\ }\href {\doibase
  10.1088/1367-2630/16/11/112002} {\bibfield  {journal} {\bibinfo  {journal}
  {New Journal of Physics}\ }\textbf {\bibinfo {volume} {16}},\ \bibinfo
  {pages} {112002} (\bibinfo {year} {2014})}\BibitemShut {NoStop}%
\bibitem [{\citenamefont {Ariando}\ \emph {et~al.}(2011)\citenamefont
  {Ariando}, \citenamefont {Wang}, \citenamefont {Baskaran}, \citenamefont
  {Liu}, \citenamefont {Huijben}, \citenamefont {Yi}, \citenamefont {Annadi},
  \citenamefont {Barman}, \citenamefont {Rusydi}, \citenamefont {Dhar},
  \citenamefont {Feng}, \citenamefont {Ding}, \citenamefont {Hilgenkamp},\ and\
  \citenamefont {Venkatesan}}]{ariandoelectronic2011}%
  \BibitemOpen
  \bibfield  {author} {\bibinfo {author} {\bibnamefont {Ariando}}, \bibinfo
  {author} {\bibfnamefont {X.}~\bibnamefont {Wang}}, \bibinfo {author}
  {\bibfnamefont {G.}~\bibnamefont {Baskaran}}, \bibinfo {author}
  {\bibfnamefont {Z.}~\bibnamefont {Liu}}, \bibinfo {author} {\bibfnamefont
  {J.}~\bibnamefont {Huijben}}, \bibinfo {author} {\bibfnamefont
  {J.}~\bibnamefont {Yi}}, \bibinfo {author} {\bibfnamefont {A.}~\bibnamefont
  {Annadi}}, \bibinfo {author} {\bibfnamefont {A.}~\bibnamefont {Barman}},
  \bibinfo {author} {\bibfnamefont {A.}~\bibnamefont {Rusydi}}, \bibinfo
  {author} {\bibfnamefont {S.}~\bibnamefont {Dhar}}, \bibinfo {author}
  {\bibfnamefont {Y.}~\bibnamefont {Feng}}, \bibinfo {author} {\bibfnamefont
  {J.}~\bibnamefont {Ding}}, \bibinfo {author} {\bibfnamefont {H.}~\bibnamefont
  {Hilgenkamp}}, \ and\ \bibinfo {author} {\bibfnamefont {T.}~\bibnamefont
  {Venkatesan}},\ }\href {\doibase 10.1038/ncomms1192} {\bibfield  {journal}
  {\bibinfo  {journal} {Nature communications}\ }\textbf {\bibinfo {volume}
  {2}},\ \bibinfo {pages} {188} (\bibinfo {year} {2011})}\BibitemShut {NoStop}%
\bibitem [{\citenamefont {Liu}\ \emph {et~al.}(2013)\citenamefont {Liu},
  \citenamefont {Li}, \citenamefont {L{{\"u}}}, \citenamefont {Huang},
  \citenamefont {Huang}, \citenamefont {Zeng}, \citenamefont {Qiu},
  \citenamefont {Huang}, \citenamefont {Annadi}, \citenamefont {Chen},
  \citenamefont {Coey}, \citenamefont {Venkatesan},\ and\ \citenamefont
  {Ariando}}]{liuorigin2013}%
  \BibitemOpen
  \bibfield  {author} {\bibinfo {author} {\bibfnamefont {Z.~Q.}\ \bibnamefont
  {Liu}}, \bibinfo {author} {\bibfnamefont {C.~J.}\ \bibnamefont {Li}},
  \bibinfo {author} {\bibfnamefont {W.~M.}\ \bibnamefont {L{{\"u}}}}, \bibinfo
  {author} {\bibfnamefont {X.~H.}\ \bibnamefont {Huang}}, \bibinfo {author}
  {\bibfnamefont {Z.}~\bibnamefont {Huang}}, \bibinfo {author} {\bibfnamefont
  {S.~W.}\ \bibnamefont {Zeng}}, \bibinfo {author} {\bibfnamefont {X.~P.}\
  \bibnamefont {Qiu}}, \bibinfo {author} {\bibfnamefont {L.~S.}\ \bibnamefont
  {Huang}}, \bibinfo {author} {\bibfnamefont {A.}~\bibnamefont {Annadi}},
  \bibinfo {author} {\bibfnamefont {J.~S.}\ \bibnamefont {Chen}}, \bibinfo
  {author} {\bibfnamefont {J.~M.~D.}\ \bibnamefont {Coey}}, \bibinfo {author}
  {\bibfnamefont {T.}~\bibnamefont {Venkatesan}}, \ and\ \bibinfo {author}
  {\bibnamefont {Ariando}},\ }\href {\doibase 10.1103/PhysRevX.3.021010}
  {\bibfield  {journal} {\bibinfo  {journal} {Phys. Rev. X}\ }\textbf {\bibinfo
  {volume} {3}} (\bibinfo {year} {2013}),\
  10.1103/PhysRevX.3.021010}\BibitemShut {NoStop}%
\bibitem [{\citenamefont {Caviglia}\ \emph
  {et~al.}(2010{\natexlab{b}})\citenamefont {Caviglia}, \citenamefont {Gabay},
  \citenamefont {Gariglio}, \citenamefont {Reyren}, \citenamefont
  {Cancellieri},\ and\ \citenamefont {Triscone}}]{cavigliatunable2010}%
  \BibitemOpen
  \bibfield  {author} {\bibinfo {author} {\bibfnamefont {A.~D.}\ \bibnamefont
  {Caviglia}}, \bibinfo {author} {\bibfnamefont {M.}~\bibnamefont {Gabay}},
  \bibinfo {author} {\bibfnamefont {S.}~\bibnamefont {Gariglio}}, \bibinfo
  {author} {\bibfnamefont {N.}~\bibnamefont {Reyren}}, \bibinfo {author}
  {\bibfnamefont {C.}~\bibnamefont {Cancellieri}}, \ and\ \bibinfo {author}
  {\bibfnamefont {J.~M.}\ \bibnamefont {Triscone}},\ }\href {\doibase
  10.1103/PhysRevLett.104.126803} {\bibfield  {journal} {\bibinfo  {journal}
  {Physical Review Letters}\ }\textbf {\bibinfo {volume} {104}} (\bibinfo
  {year} {2010}{\natexlab{b}}),\ 10.1103/PhysRevLett.104.126803}\BibitemShut
  {NoStop}%
\bibitem [{\citenamefont {F{\^e}te}\ \emph {et~al.}(2012)\citenamefont
  {F{\^e}te}, \citenamefont {Gariglio}, \citenamefont {Caviglia}, \citenamefont
  {Triscone},\ and\ \citenamefont {Gabay}}]{feterashba2012}%
  \BibitemOpen
  \bibfield  {author} {\bibinfo {author} {\bibfnamefont {A.}~\bibnamefont
  {F{\^e}te}}, \bibinfo {author} {\bibfnamefont {S.}~\bibnamefont {Gariglio}},
  \bibinfo {author} {\bibfnamefont {A.~D.}\ \bibnamefont {Caviglia}}, \bibinfo
  {author} {\bibfnamefont {J.}~\bibnamefont {Triscone}}, \ and\ \bibinfo
  {author} {\bibfnamefont {M.}~\bibnamefont {Gabay}},\ }\href {\doibase
  10.1103/PhysRevB.86.201105} {\bibfield  {journal} {\bibinfo  {journal}
  {Physical Review B}\ }\textbf {\bibinfo {volume} {86}} (\bibinfo {year}
  {2012}),\ 10.1103/PhysRevB.86.201105}\BibitemShut {NoStop}%
\bibitem [{\citenamefont {Shalom}\ \emph {et~al.}(2010)\citenamefont {Shalom},
  \citenamefont {Ron}, \citenamefont {Palevski},\ and\ \citenamefont
  {Dagan}}]{shalomshubnikovde2010}%
  \BibitemOpen
  \bibfield  {author} {\bibinfo {author} {\bibfnamefont {M.~B.}\ \bibnamefont
  {Shalom}}, \bibinfo {author} {\bibfnamefont {A.}~\bibnamefont {Ron}},
  \bibinfo {author} {\bibfnamefont {A.}~\bibnamefont {Palevski}}, \ and\
  \bibinfo {author} {\bibfnamefont {Y.}~\bibnamefont {Dagan}},\ }\href
  {\doibase 10.1103/PhysRevLett.105.206401} {\bibfield  {journal} {\bibinfo
  {journal} {Physical Review Letters}\ }\textbf {\bibinfo {volume} {105}}
  (\bibinfo {year} {2010}),\ 10.1103/PhysRevLett.105.206401}\BibitemShut
  {NoStop}%
\bibitem [{Note1()}]{Note1}%
  \BibitemOpen
  \bibinfo {note} {For sample S1, $n_{H}=1.48\times 10^{13}$ cm$^{-2}$ and
  $1.54\times 10^{12}$ cm$^{-2}<n<3.71\times 10^{12}$ cm$^{-2}$ depending on
  $B$, so that $3.98<\protect \frac {n_{H}}{n}<9.61$. For sample S2,
  $n_{H}=2.31\times 10^{13}$ cm$^{-2}$ and $3.13\times 10^{12}$
  cm$^{-2}<n<4.97\times 10^{12}$ cm$^{-2}$ depending on $B$ so that
  $4.65<\protect \frac {n_{H}}{n}<7.38$.}\BibitemShut {Stop}%
\bibitem [{\citenamefont {Xie}\ \emph {et~al.}(2014)\citenamefont {Xie},
  \citenamefont {Bell}, \citenamefont {Kim}, \citenamefont {Inoue},
  \citenamefont {Hikita},\ and\ \citenamefont {Hwang}}]{xiequantum2014}%
  \BibitemOpen
  \bibfield  {author} {\bibinfo {author} {\bibfnamefont {Y.}~\bibnamefont
  {Xie}}, \bibinfo {author} {\bibfnamefont {C.}~\bibnamefont {Bell}}, \bibinfo
  {author} {\bibfnamefont {M.}~\bibnamefont {Kim}}, \bibinfo {author}
  {\bibfnamefont {H.}~\bibnamefont {Inoue}}, \bibinfo {author} {\bibfnamefont
  {Y.}~\bibnamefont {Hikita}}, \ and\ \bibinfo {author} {\bibfnamefont {H.~Y.}\
  \bibnamefont {Hwang}},\ }\href {\doibase 10.1016/j.ssc.2014.08.006}
  {\bibfield  {journal} {\bibinfo  {journal} {Solid State Communications}\
  }\textbf {\bibinfo {volume} {197}},\ \bibinfo {pages} {25} (\bibinfo {year}
  {2014})}\BibitemShut {NoStop}%
\bibitem [{\citenamefont {Coleridge}(1991)}]{coleridgesmall-angle1991}%
  \BibitemOpen
  \bibfield  {author} {\bibinfo {author} {\bibfnamefont {P.}~\bibnamefont
  {Coleridge}},\ }\href@noop {} {\bibfield  {journal} {\bibinfo  {journal}
  {Physical review. B, Condensed matter}\ }\textbf {\bibinfo {volume} {44}},\
  \bibinfo {pages} {3793} (\bibinfo {year} {1991})}\BibitemShut {NoStop}%
\bibitem [{\citenamefont {Petrovic}\ \emph {et~al.}(2014)\citenamefont
  {Petrovic}, \citenamefont {Par{\'e}}, \citenamefont {Paudel}, \citenamefont
  {Lee}, \citenamefont {Holmes}, \citenamefont {Barnes}, \citenamefont {David},
  \citenamefont {Wu}, \citenamefont {Tsymbal},\ and\ \citenamefont
  {Panagopoulos}}]{petroviclong-range2014}%
  \BibitemOpen
  \bibfield  {author} {\bibinfo {author} {\bibfnamefont {A.~P.}\ \bibnamefont
  {Petrovic}}, \bibinfo {author} {\bibfnamefont {A.}~\bibnamefont {Par{\'e}}},
  \bibinfo {author} {\bibfnamefont {T.~R.}\ \bibnamefont {Paudel}}, \bibinfo
  {author} {\bibfnamefont {K.}~\bibnamefont {Lee}}, \bibinfo {author}
  {\bibfnamefont {S.}~\bibnamefont {Holmes}}, \bibinfo {author} {\bibfnamefont
  {C.~H.~W.}\ \bibnamefont {Barnes}}, \bibinfo {author} {\bibfnamefont
  {A.}~\bibnamefont {David}}, \bibinfo {author} {\bibfnamefont
  {T.}~\bibnamefont {Wu}}, \bibinfo {author} {\bibfnamefont {E.~Y.}\
  \bibnamefont {Tsymbal}}, \ and\ \bibinfo {author} {\bibfnamefont
  {C.}~\bibnamefont {Panagopoulos}},\ }\href {\doibase 10.1038/srep05338}
  {\bibfield  {journal} {\bibinfo  {journal} {Scientific Reports}\ }\textbf
  {\bibinfo {volume} {4}} (\bibinfo {year} {2014}),\
  10.1038/srep05338}\BibitemShut {NoStop}%
\bibitem [{\citenamefont {Jalan}\ \emph {et~al.}(2010)\citenamefont {Jalan},
  \citenamefont {Stemmer}, \citenamefont {Mack},\ and\ \citenamefont
  {Allen}}]{jalantwo-dimensional2010}%
  \BibitemOpen
  \bibfield  {author} {\bibinfo {author} {\bibfnamefont {B.}~\bibnamefont
  {Jalan}}, \bibinfo {author} {\bibfnamefont {S.}~\bibnamefont {Stemmer}},
  \bibinfo {author} {\bibfnamefont {S.}~\bibnamefont {Mack}}, \ and\ \bibinfo
  {author} {\bibfnamefont {J.~S.}\ \bibnamefont {Allen}},\ }\href {\doibase
  10.1103/PhysRevB.82.081103} {\bibfield  {journal} {\bibinfo  {journal}
  {Physical Review B}\ }\textbf {\bibinfo {volume} {82}} (\bibinfo {year}
  {2010}),\ 10.1103/PhysRevB.82.081103}\BibitemShut {NoStop}%
\bibitem [{\citenamefont {Diez}\ \emph {et~al.}(2015)\citenamefont {Diez},
  \citenamefont {Monteiro}, \citenamefont {Mattoni}, \citenamefont {Cobanera},
  \citenamefont {Hyart}, \citenamefont {Mulazimoglu}, \citenamefont {Bovenzi},
  \citenamefont {Beenakker},\ and\ \citenamefont {Caviglia}}]{diezgiant2015}%
  \BibitemOpen
  \bibfield  {author} {\bibinfo {author} {\bibfnamefont {M.}~\bibnamefont
  {Diez}}, \bibinfo {author} {\bibfnamefont {A.}~\bibnamefont {Monteiro}},
  \bibinfo {author} {\bibfnamefont {G.}~\bibnamefont {Mattoni}}, \bibinfo
  {author} {\bibfnamefont {E.}~\bibnamefont {Cobanera}}, \bibinfo {author}
  {\bibfnamefont {T.}~\bibnamefont {Hyart}}, \bibinfo {author} {\bibfnamefont
  {E.}~\bibnamefont {Mulazimoglu}}, \bibinfo {author} {\bibfnamefont
  {N.}~\bibnamefont {Bovenzi}}, \bibinfo {author} {\bibfnamefont
  {C.}~\bibnamefont {Beenakker}}, \ and\ \bibinfo {author} {\bibfnamefont
  {A.}~\bibnamefont {Caviglia}},\ }\href {\doibase
  10.1103/PhysRevLett.115.016803} {\bibfield  {journal} {\bibinfo  {journal}
  {Physical Review Letters}\ }\textbf {\bibinfo {volume} {115}} (\bibinfo
  {year} {2015}),\ 10.1103/PhysRevLett.115.016803}\BibitemShut {NoStop}%
\end{thebibliography}

%

\end{document}


\title{Supplemental Information: High field magneto-transport in two-dimensional electron gas LaAlO$_3$/SrTiO$_3$} 



\author{Ming Yang}
\email[]{ming.yang@lncmi.cnrs.fr ; phyarian@nus.edu.sg}
\affiliation{Laboratoire National des Champs Magn$\acute{e}$tiques Intenses (LNCMI-EMFL), UPR 3228, CNRS-UJF-UPS-INSA, 143 Avenue de Rangueil, 31400 Toulouse, France}

\author{Mathieu Pierre}
\affiliation{Laboratoire National des Champs Magn$\acute{e}$tiques Intenses (LNCMI-EMFL), UPR 3228, CNRS-UJF-UPS-INSA, 143 Avenue de Rangueil, 31400 Toulouse, France}

\author{Olivier Toressin}
\affiliation{Laboratoire National des Champs Magn$\acute{e}$tiques Intenses (LNCMI-EMFL), UPR 3228, CNRS-UJF-UPS-INSA, 143 Avenue de Rangueil, 31400 Toulouse, France}

\author{Michel Goiran}
\affiliation{Laboratoire National des Champs Magn$\acute{e}$tiques Intenses (LNCMI-EMFL), UPR 3228, CNRS-UJF-UPS-INSA, 143 Avenue de Rangueil, 31400 Toulouse, France}

\author{Shengwei Zeng}
\affiliation{Department of Physics and NUSNNI-Nanocore, National University of Singapore, Singapore 117411}

\author{Zhen Huang}
\affiliation{Department of Physics and NUSNNI-Nanocore, National University of Singapore, Singapore 117411}

\author{Han Kun}
\affiliation{Department of Physics and NUSNNI-Nanocore, National University of Singapore, Singapore 117411}

\author{Thirumalai Venkatesan}
\affiliation{Department of Physics and NUSNNI-Nanocore, National University of Singapore, Singapore 117411}

\author{Ariando$^*$}
\affiliation{Department of Physics and NUSNNI-Nanocore, National University of Singapore, Singapore 117411}
\email[]{phyarian@nus.edu.sg}

\author{Walter Escoffier}
\affiliation{Laboratoire National des Champs Magn$\acute{e}$tiques Intenses (LNCMI-EMFL), UPR 3228, CNRS-UJF-UPS-INSA, 143 Avenue de Rangueil, 31400 Toulouse, France}

\author{Michael Coey}
\affiliation{School of Physics and CRANN, Trinity College, Dublin, Ireland}
\affiliation{Department of Physics and NUSNNI-Nanocore, National University of Singapore, Singapore 117411}

\date{\today}

\maketitle 

\section{Transport measurement in high pulsed magnetic field}
\begin{figure} [ht]
\includegraphics[width=\linewidth]{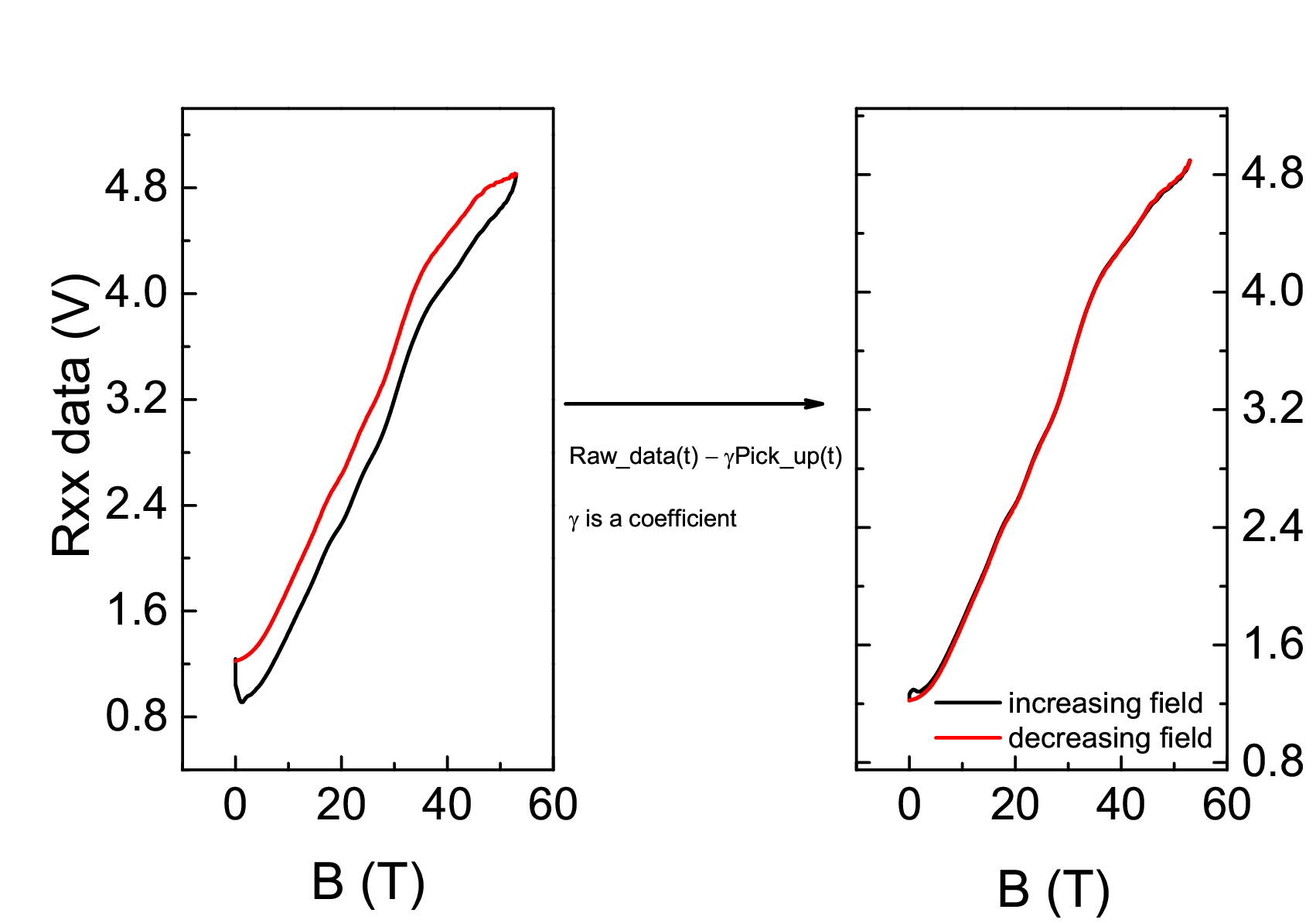}
\caption{\label{fig.S1} Raw data as a function of magnetic field before(a) and after(b) compensation of the induced voltage in DC measurement scheme.}
\end{figure}

High-field magneto-transport measurements are performed using the pulsed magnetic field technique. The magnetic field impulsion reaches its maximum value of 55T within 60ms, and then decays exponentially down to 0T within a time scale of roughly 300ms. A fast data acquisition system (up to 2MHz) is used to record the voltage difference between the electrodes of the sample during the full duration of the magnetic field pulse. The current passing through the sample is simultaneously recorded and the ratio of voltage over the current provides the magneto-resistance. Contrary to the static magnet technique, time-varying magnetic field introduces spurious additional voltages $\delta V$ at the sample's electrodes that is proportional to the time-derivative magnetic flux $\Phi$ enclosed by the wiring loop ($\delta V=-\frac{d\Phi}{dt}$). Despite special care in wiring geometry to minimize the magnetic flux, the induced-voltage is always present and removed from the signal using two methods:
\begin{itemize}
\item In DC measurement (e.g. the current passing through the sample is steady), a fraction of the pure induced voltage measured by a pick-up coil in the close vicinity of the sample is subtracted to the raw recorded signal. The fraction (e.g. the ratio between the total area of the pick-up coil and the area of the sample's wiring loop) is chosen so that the magneto-resistance is similar both when the magnetic field increases and decreases (see figure \ref{fig.S1}). This technique provides reliable measurements, however the signal-to-noise ratio is limited by the maximum current injected in the device.
\item In AC measurement (e.g. the current passing through the sample is sinusoidal with frequency $f_1>5$ KHz), the parasitic induced voltage is naturally removed during demodulation since the natural frequency range of the pulsed field $0<f_0<1$KHz is much lower than the current modulation frequency $f_1$. For demodulation of the raw signal, we used a home-made numerical lock-in. This technique provides a higher signal-to-noise ratio and allows the measurement of small magneto-resistance oscillations. On the other hand, the back-ground signal is often distorted (as compared to DC measurements) because of unwanted capacitances/inductances associated with the measurement electrical circuit, which become relevant at high frequency. It has been checked that the AC/DC measurement schemes provides almost similar results when dealing with the small magneto-resistance oscillations (neglecting extra-noise in DC measurement).
\end{itemize}
In the main article, magneto-resistance oscillations (SdHO) are obtained using the AC measurement technique while the full magneto-resistance are recorded using the DC measurement technique.

\section{Two-fluid model with magnetic field dependent carrier density}
\begin{figure} [h]
\includegraphics[width=\linewidth]{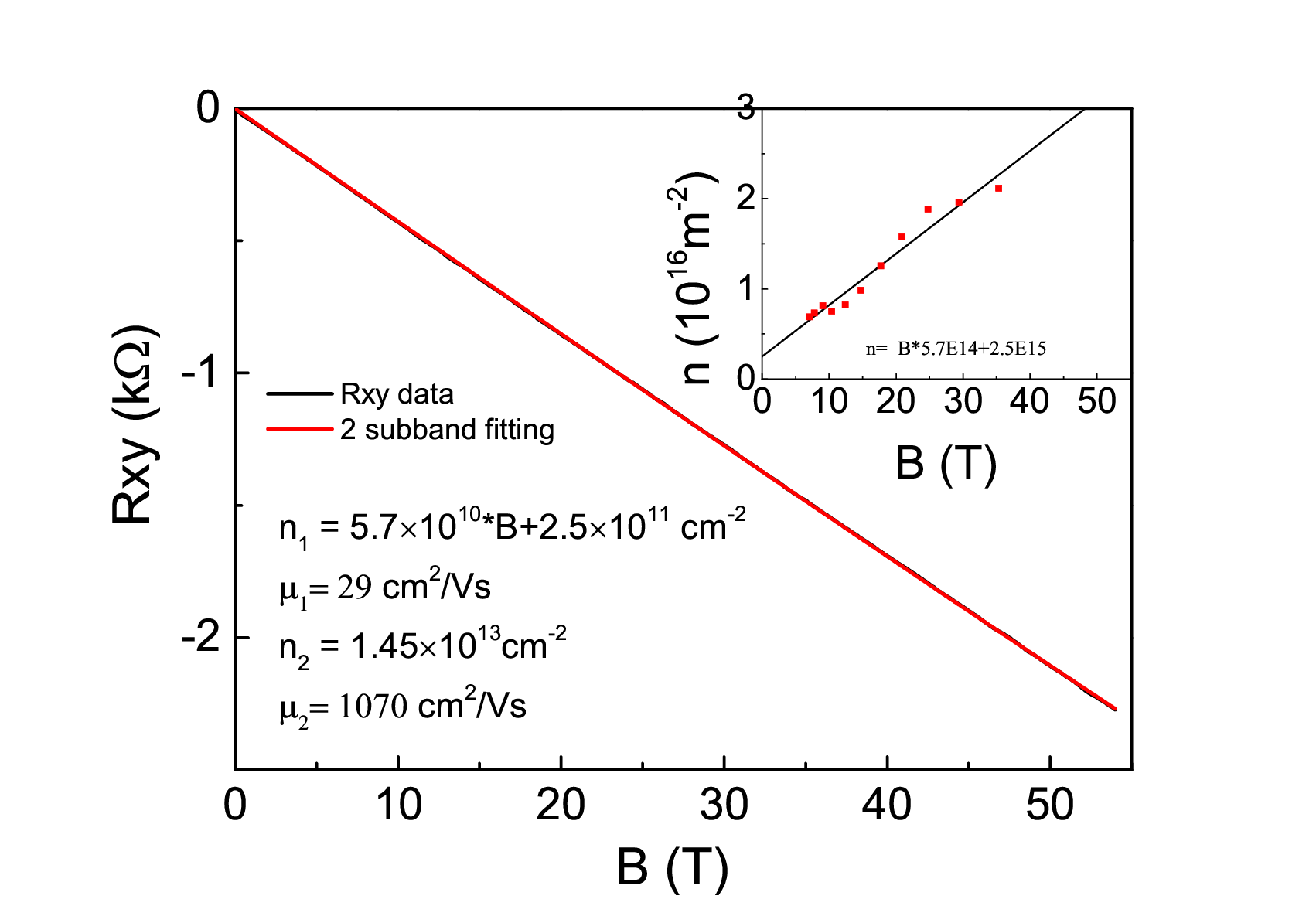}
\caption{\label{fig.S2} Experimental Hall resistance data (black) and two-fluid model fitting with a B-dependent carrier density $n_1$ (red). The best free parameters are $n_2=1.45\times 10^{13}$cm$^{-2}$, $\mu_1$=29cm$^2$/Vs and $\mu_2$=1070cm$^2$/Vs. The insert shows the carrier density $n_1(B)$ as a function of magnetic field extracted from the Onsager relation.}
\end{figure}

Since a period shift of the SdH oscillations is observed when plotted versus inverse magnetic field, we assume that the carrier density could be magnetic field dependent. For selected values of $B$, the carrier density is therefore extracted from two consecutive peaks/valleys of the oscillations using the Onsager relation. The evolution of the carrier density $n_1(B)$ is then interpolated over the full magnetic field range as shown in the insert of figure \ref{fig.S2}. $n_1(B)$ is introduced in the two-fluid model in order to fit the Hall resistance. As shown in figure \ref{fig.S2}, the linear Hall resistance is recovered but the free fitting parameters ($n_2$, $\mu_1$ and $\mu_2$) are un-physical (notice the very low mobility $\mu_1$=29cm$^2$/Vs for carriers giving rise to SdH oscillations). Therefore, the Rashba spin-orbit coupling hypothesis is favored in order to explain the period shift of the SdH oscillations, rather than a magnetic field dependent carrier density.

\section{Negative magneto-resistance in parallel magnetic field at 20K}
\begin{figure} [h]
\includegraphics[width=0.6\linewidth]{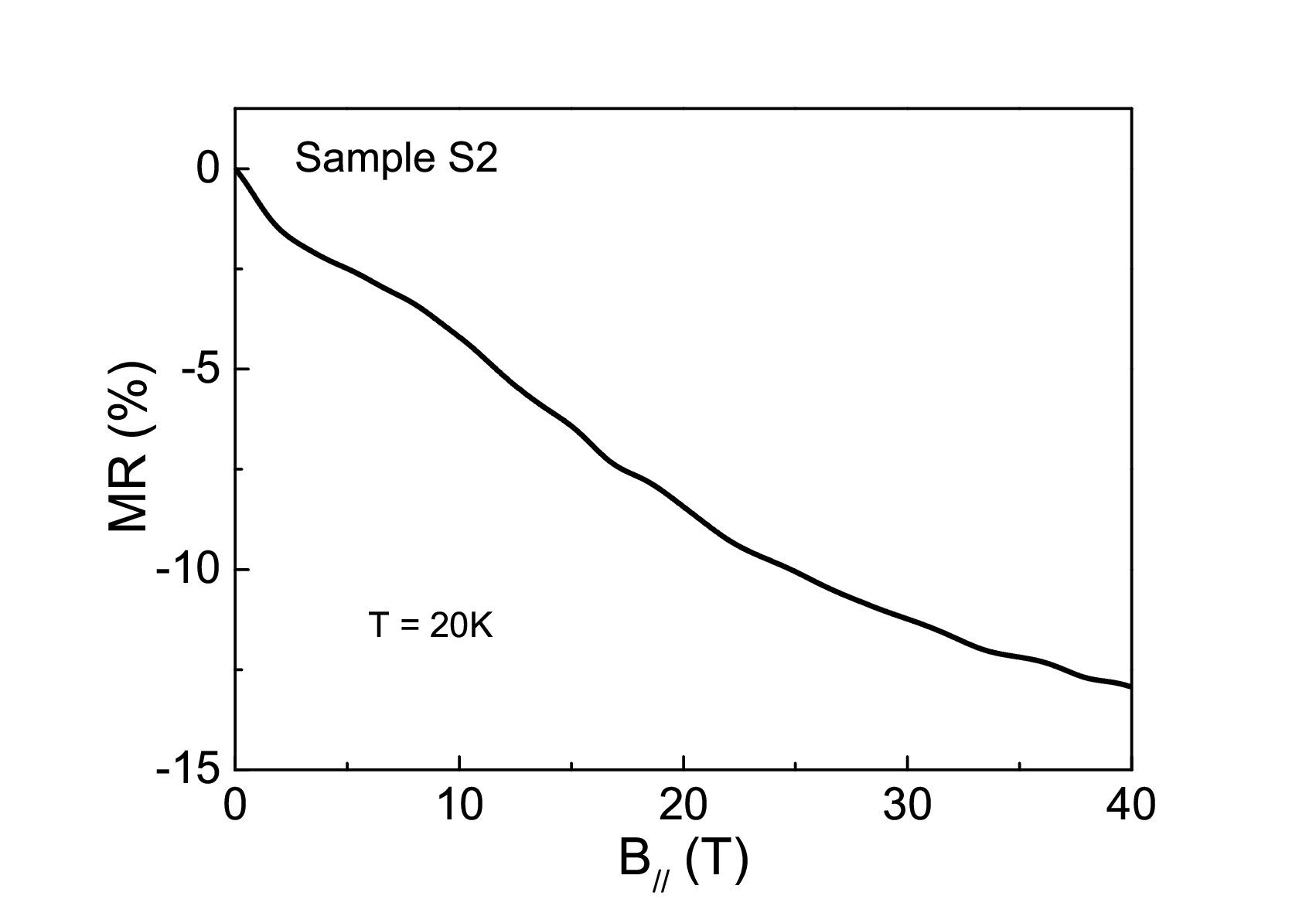}
\caption{\label{fig.S3} Negative magneto-resistance at 20K under a parallel field.}
\end{figure}

We measured the magneto-resistance at 20K under a parallel field up to 40T. A 15\% negative MR can be observed in figure \ref{fig.S3}. The persistence of the negative MR at 20K indicates that the Kondo model is unlikely to account for the negative MR in our case.

\section{Fit of SdHO in the presence of spin-orbit coupling}
In the following figures, the relative influence of the fitting parameters is investigated. The figures presented in the main text correspond to the best fitting parameters.
\begin{figure} [h]
\includegraphics[width=\linewidth]{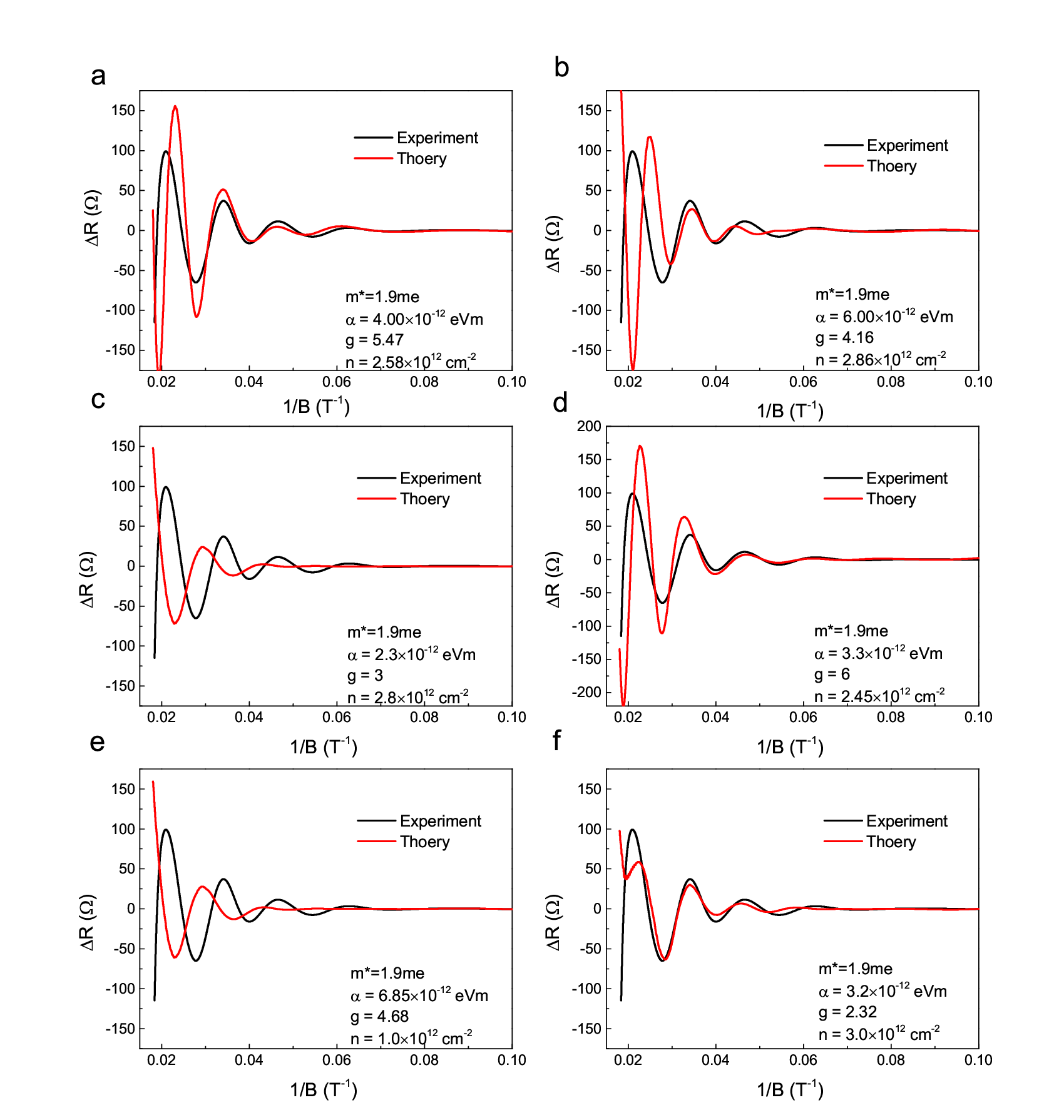}
\caption{\label{fig.S4} For all figures, the effective mass is constrained to the value $m^*=1.9 m_e$ extracted from SdH analysis. (a) and (b) The S.O. coupling strength is constrained above and below the best value ($5.45 \times10^{-12}$ eVm) while the g-factor and carrier density are free parameters, (c) and (d) The g-factor is constrained above and below the best value ($g=5$) while the S.O. coupling strength and carrier density are free parameters, (e) and (f) The carrier density is constrained above and below the best value ($n=1,65\times10^{12} cm^{-2}$) while the S.O. coupling strength and g-factor are free parameters}
\end{figure}

\section{Background removal}
In the raw data, the amplitude of the SdH oscillations is small, overwhelmed by a large monotonous background. The background is extracted by manually pinpointing the middle between two neighboring peaks or valleys. The full background data is built using a cubic-spline interpolation routine. The figure \ref{fig.S5} below show an example of background extraction for sample S1.
\begin{figure} [h]
\includegraphics[width=\linewidth]{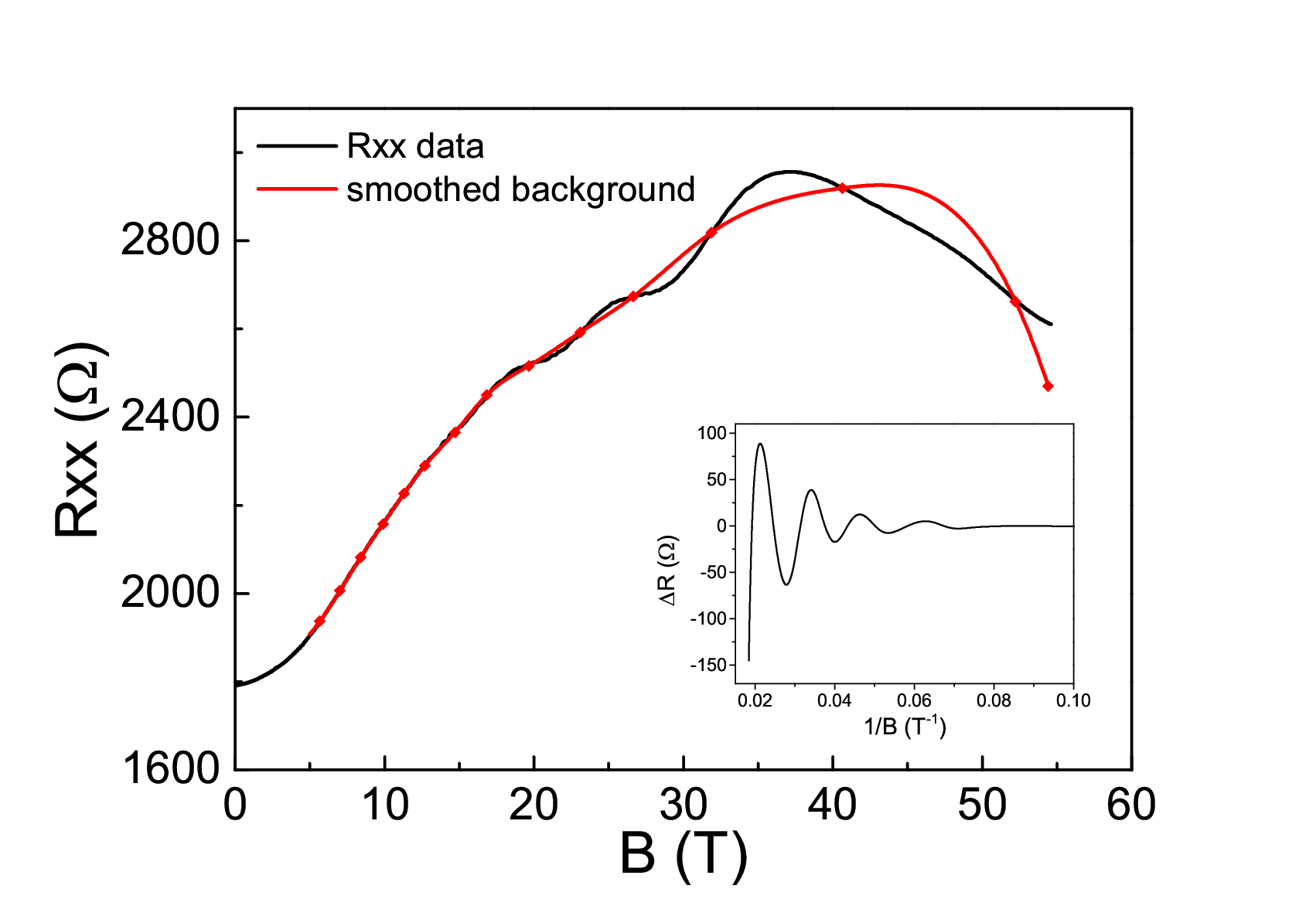}
\caption{\label{fig.S5} Raw signal and interpolated background signal for sample S1}
\end{figure}